# Doping of Cr in Graphene Using Electron Beam Manipulation for Functional Defect Engineering

*Ondrej Dyck,[1][*] Mina Yoon,[1] Lizhi Zhang,[2] Andrew R. Lupini,[1] Jacob L. Swett,[3] Stephen Jesse[1]*

[1] *Center for Nanophase Materials Sciences, Oak Ridge National Laboratory, Oak Ridge, TN*

[2] *Department of Physics and Astronomy, University of Tennessee, Knoxville, TN*

[3] *Department of Materials, University of Oxford, Oxford OX1 3PH, UK*

[*] *Corresponding Author E-mail: dyckoe@ornl.gov*

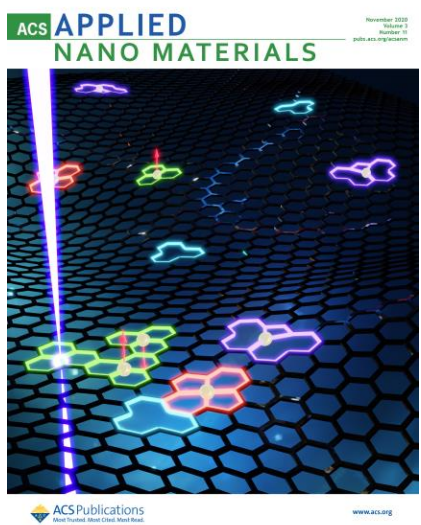

## Abstract

Chromium atoms in graphene have been proposed to exhibit magnetic moments and spin-selective conducting states depending on the local bonding geometry within the graphene structure, which could lead to interesting applications in spintronics. Despite this interest, there are few direct experimental reports of Cr dopants in graphene even though it is theorized to be stable. Here, we demonstrate the introduction of single Cr dopant atoms into the graphene lattice and onto graphene edges through the controlled use of a focused electron beam in a scanning transmission electron microscope. We show local control of doping locations and when coupled with targeted *in situ* milling during scanning of the e-beam, these strategies demonstrate an important component of the fabrication of tailored nanostructured devices in the electron microscope. The approach is validated with first principles calculations to understand synthesis pathways during fabrication and reveal the energetics and local properties of Cr atoms embedded in graphene; e.g., Cr doping can convert graphene into a magnetic and semiconducting material, which suggests Cr-doped graphene can be used as a building block for potential electronic devices and a means to construct them.

**Introduction**

Breakthroughs that enable the design and synthesis of material systems with specific properties at targeted locations have driven the information age. The foremost examples are optical and electron lithography that enabled targeted, highly precise doping and thin film growth/deposition over large areas. Progression to smaller length scales, which is a key to the success of this paradigm, is running up against a fundamental transition. Materials at sufficiently small length scales switch from exhibiting continuum behavior to exhibiting properties dictated by the underlying discretized atomic nature of matter and the behavior of only a handful of atoms. Some existing foundational device architectures and fabrication approaches might not make this transition; however, new opportunities are being created in which a small number of atoms, precisely arranged within a solid substrate, can be designed and fabricated with novel, useful, and addressable functionality.

Building at the discretized level of matter allows many advantages for reliability and repeatability.[1] This is extremely useful for devices in general, but even more so for emerging quantum devices that rely on entanglement where even slight differences between neighboring quantum states, resonances, etc. (which can easily happen in larger scale transmon type qubits, for example) can dramatically degrade coherence lifetimes and ultimately will severely limit the scale of quantum computation. If the quantum centers responsible for calculations are limited to a small number of atoms (artificial molecules) that are effectively identical, then their behaviors can be far more consistent and coherent. Furthermore, if the spacing between quantum centers can be set by a specific number of unit cells within a crystal lattice, then their interactions will be far more regular and predictable as well. It is therefore desirable to develop pathways for engineering functional building blocks that can be positioned with high spatial resolution and exhibit useful physical properties that can be technologically harnessed.

Here, we demonstrate spatially controllable introduction of individual Cr dopants into a suspended two-dimensional (2D) graphene lattice within a range of vacancy configurations to tune local electronic and magnetic properties. Studies have shown that doping graphene with Cr can confer unique properties for designing next-generation quantum and other devices, inducing half metallicity, 100% spin polarization at the Fermi energy, a small band gap, and a magnetic moment.[2-4] Graphene has received much attention since its discovery, where it was mechanically isolated from bulk graphite.[5] Since then, successful growth of large-area, high-quality graphene[6] has fueled the collective scientific imagination, with many potential applications[7-9] including graphene nanoribbon field-effect transistors,[10-12] the half-integer quantum hall effect,[13, 14] spintronic properties,[8, 15, 16] and quantum qubits,[17-20] all of which sensitively depend on the precise graphene geometry. For example, zig-zag and arm-chair terminated graphene nanoribbons show a non-linear variation of the band structure, spin density, and molecular orbitals as a function of nanoribbon width.[21] It is well known that doping graphene can change its electronic structure and several theoretical investigations have taken steps to explore these possibilities.[22-24] Despite the prediction of interesting properties, there has been little experimental work performed on Cr-doped graphene.

Previous reports of successful transition metal doping of graphene include: using thermal evaporation and subsequent incorporation at existing defects;[25] and deliberately creating a large number of defects through plasma treatment or by wide area electron beam irradiation followed by heating to cause metal atoms to diffuse.[26] Both of those methods can potentially be scaled to produce large areas of doped graphene but, as described, they do not provide precise atomic scale control of the location and number of dopants incorporated. To achieve controllable creation of atomic scale building blocks for the fabrication of novel functionality and devices, it is necessary to develop techniques that can be applied in a highly localized manner.

To achieve high spatial precision in material synthesis of Cr-doped graphene described here, we use the focused electron beam (e-beam) of a scanning transmission electron microscope (STEM) to create vacancies, insert dopants, and evolve the local structure. The utilization of STEM for atomic assembly is a fairly recent advancement that has been demonstrated as a viable approach for atomic-scale material sculpting[27, 28] and manipulation.[29, 30] It was recently demonstrated that the e-beam could be used to controllably attach foreign atoms to defect sites in graphene.[31-33] The process involves e-beam-induced ejection of C atoms from a desired location followed by e-beam-induced sputtering of nearby dopants. The sputtered atoms then attach to the defect sites. Such tasks were first performed with Si atoms, which were convenient since they are often found in the surface contamination on graphene. This strategy was extended to include the insertion of Pt atoms and Pt-Si heteroatomic clusters in graphene.[34] In the present study, Cr atoms are introduced onto the graphene surface using e-beam evaporation *ex situ* and then using the STEM e-beam to scatter the Cr atoms into lattice defects created *in situ*. This "atomic-scale material synthesis laboratory" provides the ability to visualize the evolution of structures in real time. This allows one to quickly and experimentally ascertain the most likely and stable configurations and to directly translate the observed atomic positions to first principles modeling to predict electronic and magnetic properties. Density functional theory (DFT) simulations were informed by the experimentally observed structures to better understand the effects of Cr-doping on the electronic and magnetic structures of graphene. The results presented lay the groundwork for directed synthesis with an ultimate goal of enabling a translation in material synthesis for the next generation of materials and devices for quantum information science applications.

**Results and discussion**

To permit *in situ* STEM e-beam driven insertion of Cr atoms as dopants in the graphene lattice, appropriate graphene samples were prepared as described in the supplemental materials. Briefly, CVD-grown graphene was transferred onto microporous SiN STEM grids purchased from TEMwindows.com and Cr was evaporated onto the surface creating Cr nanoparticles and single dispersed atoms that provide a ubiquitous source of Cr. A schematic of the sample preparation process is shown in Figure 1. It should be noted that the common contaminants that adhere to the graphene surface are also present with the intentionally deposited Cr atoms.[35] This is the source of the observed Si in the present work.

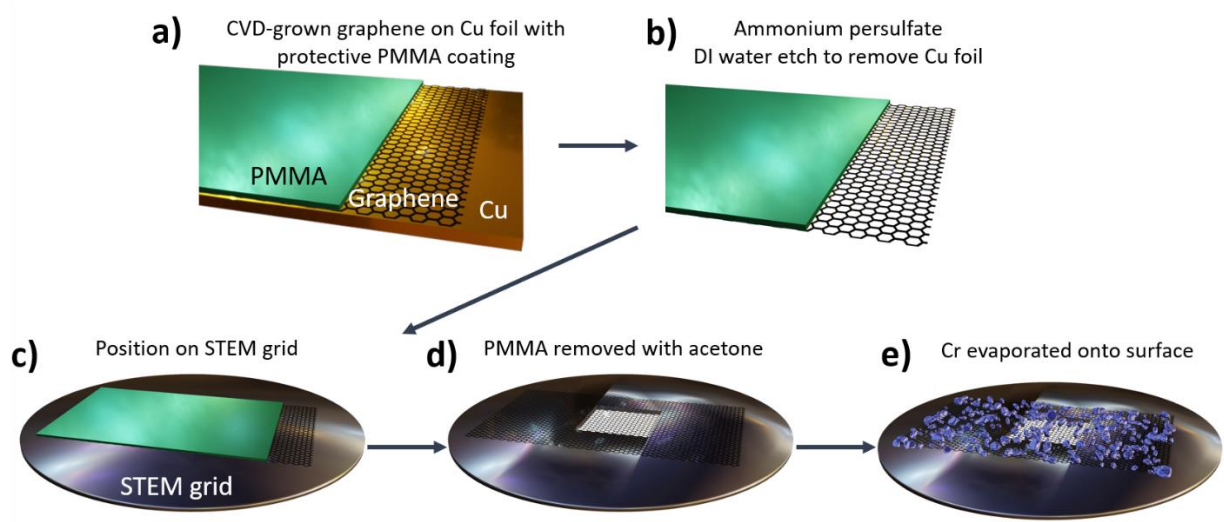

**Figure 1 Schematic illustrating sample preparation workflow**. a) Graphene was grown on Cu foil and coated with poly(methyl methacrylate) (PMMA). b) An ammonium persulfate/DI water bath was used to etch away the Cu foil.

c) The PMMA/graphene stack was put on a STEM grid. d) Acetone was used to remove the PMMA. e) Cr was evaporated onto the graphene surface.

Electron energy loss spectroscopy (EELS) was used to verify the identity of each element and is shown in the supporting materials. We selected an area on the graphene where a Cr nanoparticle was ~1 nm away from a pristine area on the graphene. The 100 kV e-beam was positioned on the edge of the Cr nanoparticle and was moved manually back and forth from the edge of the nanoparticle onto the pristine are of the graphene lattice. This process, detailed in Figure 2, was adapted from previous results describing the *in situ* insertion of Si dopants into a graphene lattice where e-beam-induced defect sites trap atoms scattered by the e-beam across the surface.[31, 32] Figure 2a shows the initial configuration where the arrow indicates the beam dragging direction. Figure 2b shows the first notable structural change where a Cr dopant has been scattered into a defective region of the graphene lattice. Figure 2c and 1d show the subsequent configurations; in Figure 2d, a small hole appears in the graphene lattice with a Cr and a Si atom attached to edge of the hole adjacent to each other. To prevent continued damage to the graphene, e-beam dragging in this region was stopped. A larger area was scanned to lower the electron dose per area and to sputter the surrounding C atoms to heal the hole. This is illustrated in Figure 2e where the hole has been healed and additional Cr atoms have been locally incorporated into the graphene. A STEM video (see supplemental materials) captured the evolution of the graphene/Cr/Si system during local e-beam irradiation. Figure 2f-h show several video frames where a few more dopants are captured in 1f and the appearance and growth of a new hole in shown in 1g and 1h. Notably, we also observe that a heteroatomic chain of Cr and Si atoms forms at the edge of this new hole in the graphene.

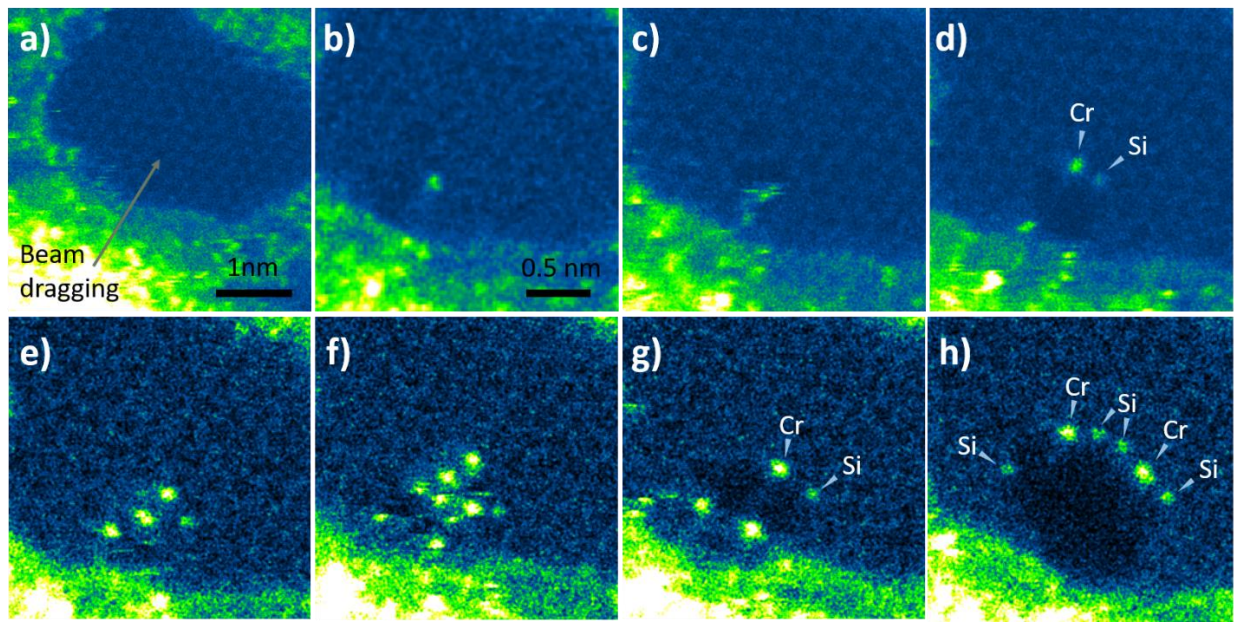

**Figure 2 Using the e-beam to scatter Cr and Si dopants onto the graphene lattice and onto the edge of an e-beam-induced hole.** a) Medium angle annular dark field (MAADF)-STEM image of the initial sample state where a pristine area of graphene is located next to a Cr nanoparticle (bright area in lower left corner). An e-beam dragging technique was used to scatter atoms from the Cr nanoparticle and surrounding amorphous material onto the graphene surface as well as eject C atoms from the graphene lattice. b) Cr atom scattered onto the edge of an e-beam-induced nanopore. c) Structure evolves under the e-beam as evidenced by streaking of the Cr atom. d) Si atom scattered onto the edge of the nanopore next to the Cr atom. e) Nanopore heals and incorporates additional Cr atoms. f)-h) Continued structural evolution under the e-beam. Images cleaned by applying the smooth function in Fiji followed by gaussian blur. Artificially colored using the green-fire-blue LUT in Fiji.[36]

We note that the e-beam is playing a dual role here, namely, it is being used both to scatter atoms across the graphene surface from the source particle, as well as generate defect sites to which the source atoms can attach. For the purposes of this study, this requires the source material to be very close to the doping location on the surface of the sample. Further improvement in defect site positioning could be accomplished by delivering a continuous flow of source atoms across the graphene surface so that the e-beam is only needed to create defect sites.

With a method to locally attach Cr dopants to defect sites in graphene next we begin to explore what structural transitions can be driven by the e-beam. Previous experiments[28, 31, 32, 37-40] demonstrated that it is possible to deterministically move single Si dopants through the graphene lattice, leveraging a bond inversion process driven by a 60 kV e-beam. For the movement of Si this involves positioning the e-beam on a C atom adjacent to the Si dopant to trigger the bond inversion process. The strategy used to discover this process involved irradiating an area with a dopant, documenting the evolution of the system, and finally modelling the atomic behavior theoretically to understand where to position the e-beam to achieve the desired transformation. Extending this or a similar capability to other dopant elements would represent a significant advance and open the possibility of atomic scale assembly of structures involving different atomic species.[29] By using observed structures, we can turn to theoretical treatments to determine if they exhibit valuable properties. To explore the structural transformations in the Cr-graphene system, we document the evolution of a 3-fold C-coordinated Cr dopant under 100 kV e-beam irradiation, Figure 3, which was inserted into the lattice by scanning the e-beam over the source material and the pristine graphene. By using the 100 kV e-beam for this experiment, we are allowing a wide range of structural transformations to take place up to and including complete ejection of C atoms from the lattice. The Cr-dopant begins by occupying a single vacancy in the graphene lattice, as shown in Figure 3a; a magnified view is shown in Figure 3b where the Cr-dopant and surrounding C atoms are marked by colored dots. The first transformation observed during this experiment was the ejection of two C atoms adjacent to the Cr dopant (circled in Figure 3b), which results in the Cr dopant atom moving slightly to fill the tri-vacancy, as shown in Figure 3c and 2d. The next transition was the capture of a C atom and conversion of the tri-vacancy to a di-vacancy. The position where the C atom was added is marked on Figure 3d by a circle and the C atom is indicated by an arrow in

Figure 3f. Next, the C atom circled in Figure 3f was ejected converting the structure back to a tri-vacancy. The subsequent transformations observed (Figure 3i-p) become more complex and involved the creation of an increasing number of clustered defects and ejection of additional C atoms. While more defects are observed in the STEM images (away from the Cr atom) we highlight the positions of all the C atoms forming the defect cluster up to the first observed 6-member graphene rings. The expansion of the defect area surrounding the Cr atom is evident from Figures 2i-p. While the Cr atom was observed to change positions with respect to the lattice (Figure 3a-h), this was due to the sequential ejection, restructuring and later recapture of C atoms; Figure 3i-p show this was due to the restructuring and accumulation of surrounding defects. In contrast, the movement mechanism for Si in graphene involves partial ejection and immediate recapture of C atoms,[28] which leave the lattice symmetry unchanged during manipulation. In the present experiment, the surrounding lattice is changed at every stage/step, making it difficult to determine a repeatable movement process from the e-beam-induced structural evolution observed.

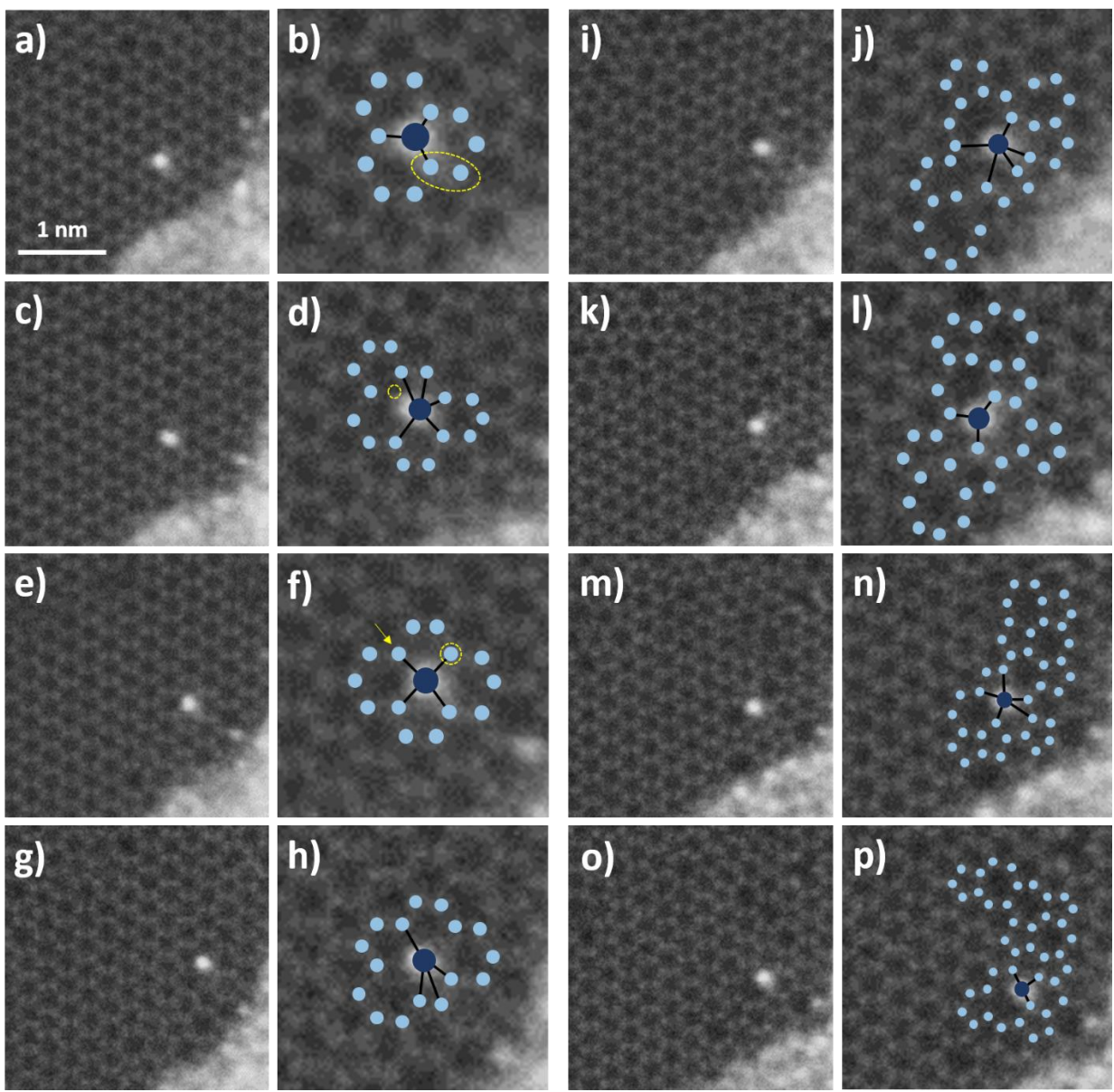

**Figure 3 Structural evolution under a 100 kV e-beam.** a) Cr dopant introduced into a graphene vacancy by scanning the e-beam over both source material and the pristine graphene lattice. b) Magnified view of the defect with Cr atom and surrounding C atoms marked by closed circles. Dotted circle shows two C atoms that were ejected from the graphene lattice to produce the structure shown in c)-d). The Cr atom in c) and d) occupies a tri-vacancy and dotted circle in d) marks location where a C atom is recaptured, shown in e)-f). f) Recaptured C atom marked by the arrow. The Cr atom occupies a di-vacancy and is 4-fold coordinated. A circle marks the C atom that was ejected to produce the structure shown in g)-h). The Cr atom occupies a tri-vacancy again. i)-p) show the continued evolution of the structure where an expanding cluster of e-beam-induced defects surrounds the Cr atom.

Based on these experimental observations of the Cr-dopant/graphene structures, we performed first-principles DFT calculations to gain insight into the energetic, electronic, and magnetic properties of these structures (see Methods section) and determine whether emergent physical properties exist which may be of interest technologically. We consider a Cr-dopant that occupies a single (mono)vacancy, di-vacancy, and tri-vacancy in graphene (consisting of 50 atoms without vacancies) compared with the vacancy structures alone. Figure 4 shows the spin-polarized density of states (DOS), where the black and red lines represent the total DOS and the states projected onto the Cr dopants (left panel) or onto the atoms near the defect sites (right panel), respectively. All the Cr-doped structures exhibit a significant contribution of the Cr atom to an unbalanced spin polarization at several energies in a wide energy range (~-3eV to ~+4eV). The divacancy graphene with no Cr dopant exhibits a balanced spin up and down contribution. The tri-vacancy has a minor contribution, and the monovacancy has the most noticeable spin unbalanced state in a narrow energy range around the Fermi level.

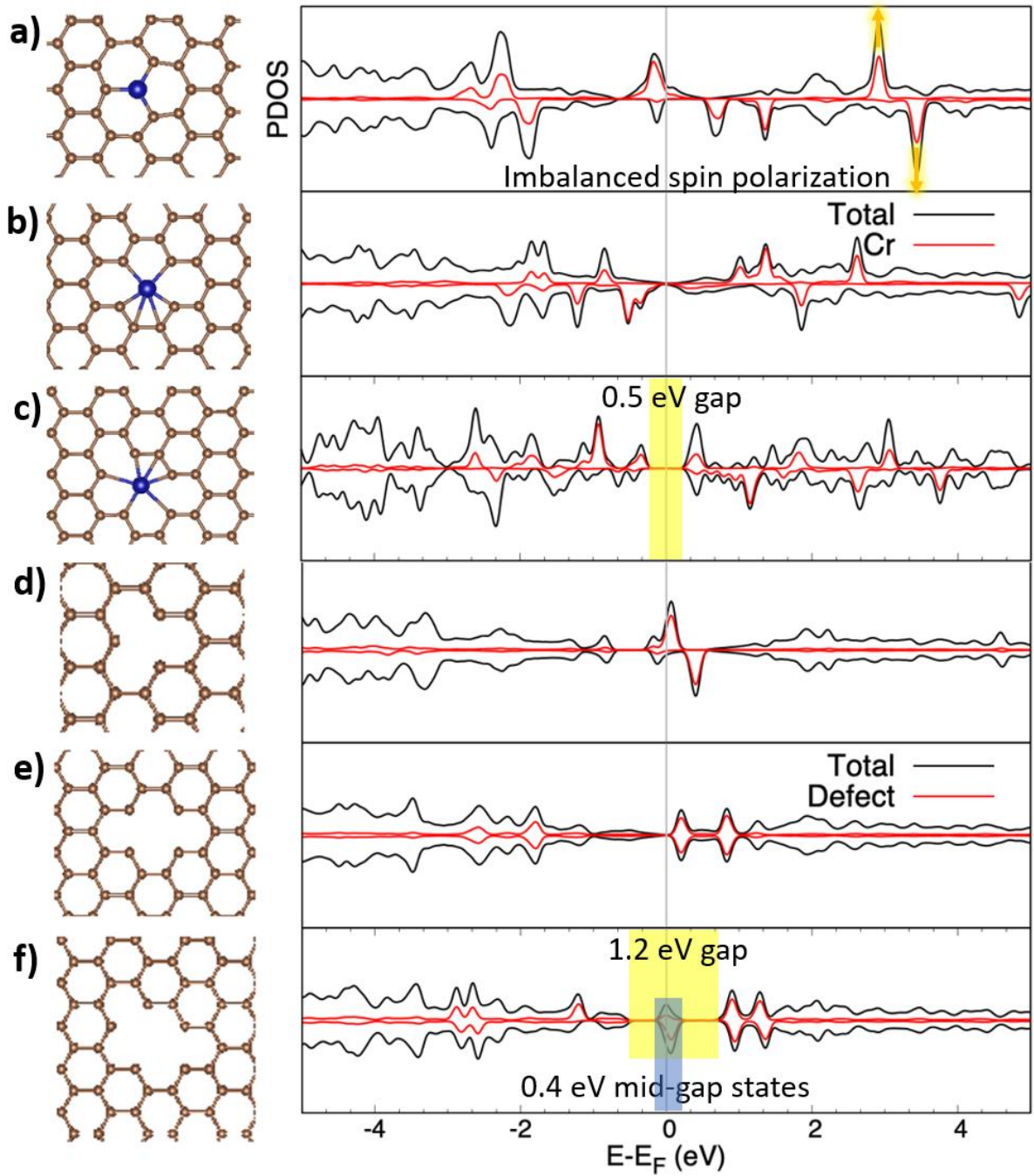

**Figure 4** First-principles modeling of Cr-embedded graphene defects with mono (a), di (b), and tri (c) carbon vacancies, where blue and brown balls represent Cr and C atoms, respectively. The left panel shows the optimized metal-embedded structures and their atomic site projected density of states (PDOS); they are compared with those of graphene with vacancies in the right panel. Total DOS are presented in black lines and the red ones are projected onto either Cr atoms (left panel) or atoms near defect sites (right panel). Energy scale in the PDOS is presented with respect to Fermi level ($E_F$).

A C monovacancy creates three C atoms with missing bonds in the xy plane, resulting in defect states consisting of $p_x$ and $p_y$ orbitals shifted up near the Fermi level that have spin-up and down states separated by ~0.5eV with a total magnetic moment of 0.5$\mu_B$. The defect destroys the local π conjugated graphene network from $p_z$ orbitals, which opens a sizable electronic gap ~0.5eV below the Fermi level. As the defect becomes occupied by a Cr atom, charge transfer from the Cr to C occupies some of the defect states and thus, they are shifted below the Fermi level. A strong hybridization between the $d_{z2}$ orbital of Cr and $p_z$ orbitals of the C atoms results in a peak just below the Fermi level, where the defect C state in the minority spin is aligned with the majority spin state of Cr located below the Fermi level, giving a total magnetic moment of 2.0 $\mu_B$. However, at ~3eV above the Fermi energy we observe a much larger unbalanced spin polarization in the Cr-doped graphene with spin up and spin down components separated by ~0.5eV, which does not appear in the non-occupied defect. This property allows for the possibility of electronically controlled magnetic switching. For the Cr atom occupying a divacancy we observe a low DOS as the local π conjugation is further destroyed by creating C vacancies. The spin up and spin down components near the Fermi level (within ~2eV) are exclusively from the hybridization between *d* orbitals of Cr and *p* orbitals of C atoms. The naked divacancy has four under-coordinated C atoms arranged in a symmetric way; thus, it has completely balanced majority and minority spin components. As we create one more C vacancy, the π conjugation of graphene is completely destroyed and the tri-vacancy has a wide range of gap states (~1.5eV) with a peak at the Fermi level; the peak consists of the majority spin state from $p_z$ orbitals and the minority state hybridized between the *p* orbitals. As the Cr atom occupies the tri-vacancy, charge transfer from the metal to carbon shifts down the defect states as well as the gap state and a larger energy gap opens at the Fermi energy of 0.5 eV. This is distinct from the naked tri-vacancy where a ~1.2 eV gap exists with a ~0.4 eV wide defect state in the middle of the gap. The presence of the Cr atom decreases the gap width but suppresses the mid-gap states. It should be noted that these properties are based on the specific details of the simulated Cr-doped graphene with an infinitely repeating, idealized crystal structure. Thus, it does not represent the experimental case exactly but instead serves to illustrate the promising aspects of the Cr-doped graphene system.

Another question to address is what structures might be possible to fabricate. Atomic scale fabrication is subject to significant limitations imposed by chemistry and bonding[41] and it is not obvious what structures will be stable. Using a theoretical approach is difficult because of the large number of possible configurations as the number and types of atoms increases. An additional difficulty is that the energetic pathways from one structure to the next also need to be mapped out if the intent is to understand what steps should be taken for fabrication. Experimentally, this problem can be handled more directly by irradiating the sample and documenting what structures are observed. These will be the most likely ground state structures and significantly narrows the configuration space for theory. We used this approach to explore heteroatomic arrangements involving Cr, Si, and C, generated by exposing the graphene and surrounding source material to the 100 kV e-beam. The defects created in the graphene become attachment sites for atoms sputtered from the surrounding source material. Continued e-beam exposure causes significant restructuring of the defective graphene lattice where we observe the formation of various Si-Cr heteroatomic clusters. These experimentally observed structures were used to guide the theoretical investigations.

Figure 5 shows MAADF-STEM images of the experimentally created heteroatomic structures. The insets in 4a and 4b show structural models based on the atom positions determined from the STEM images. Figures 4c-f show additional examples. These observations indicate that heteroatomic Si-Cr structures can be inserted into a graphene lattice. This capability, combined with other recent advances in the field,[42] collectively represent an initial step toward constructing locally defined structures that exhibit desired functionality. Videos of the dynamic rearrangement of these structures under the 100 kV e-beam are provided in the supplementary materials.

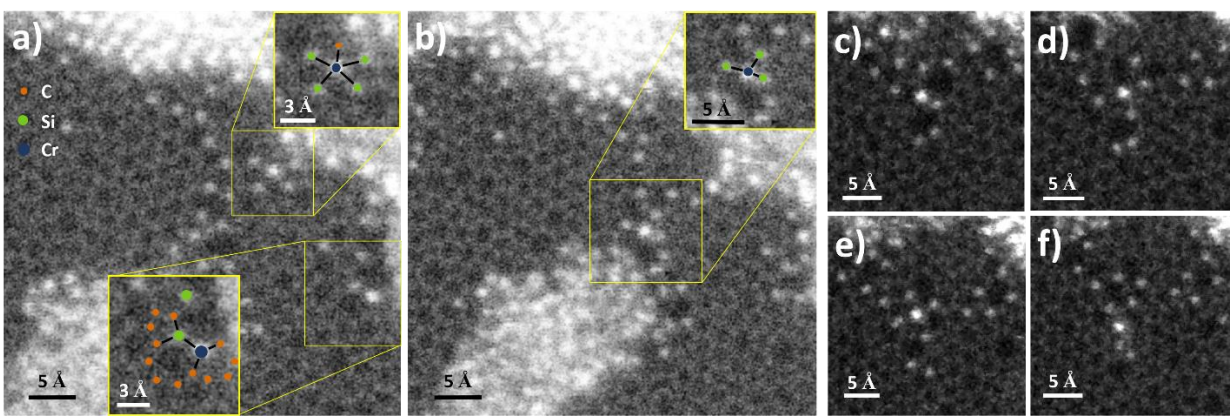

**Figure 5 Summary of experimentally observed heteroatomic structures.** In a) and b) boxes highlight locations where we observe Si-Cr heteroatom clusters. Overlays highlight atomic structures in the immediate neighborhood around the Cr atom. c)-f) show additional examples of Cr-Si heteroatomic clusters in the defective graphene lattice.

To more clearly understand the energetic processes involved, DFT simulations were performed on a few representative configurations. Experimentally, we observed that the creation of defect clusters seems to be a two-step process beginning with the creation of point defects and followed by the subsequent attachment of foreign atoms to these defect sites. Based on these observations we attempt to model a plausible series of alterations that should result in a similar structure.

The DFT calculations evaluated the energetics to create Stone-Wales (SW) defects and the atomic vacancies associated with Si and Cr dopants. We modeled doped-SW defects using a supercell of 180 carbon atoms including one SW defect with one or two carbon atoms replaced by Cr or Si atoms, Figure 6a-e, to explore the energetics required to drive point defect creation and subsequent attachment of Cr and/or Si dopants to the defect site.

Alternatively, several C atoms could be ejected from the lattice prior to the attachment of dopant atoms. The second simulation series, Figure 6a and 5f-i, begins with a reconstructed graphene nanopore with six atomic vacancies (176 carbon atoms in the supercell) to examine the role of passivating Si and Cr atoms. The atomic positions were inspired by the experimental observations (Figure 4) and fully optimized within the DFT calculations. Here the formation enthalpy (DE) is defined as $\Delta E = E_{total} - (n_C E_{graphene} - n_{Si} \mu_{Cr} - n_{Cr} \mu_{Cr})$, where $E_{total}$ is the total energy of the system consisting of defects/vacancies and dopants, $E_{graphene}$ is the energy of pristine graphene per atom, $m_{Cr}$ and $m_{Si}$ are the chemical potential of Cr and Si atoms, respectively, in the gas phases, and $n_C$, $n_{Si}$, $n_{Cr}$ represent the number of C, Si, and Cr atoms in the system, respectively. The chemical potentials of the gas species are taken from the experimental data.[43]

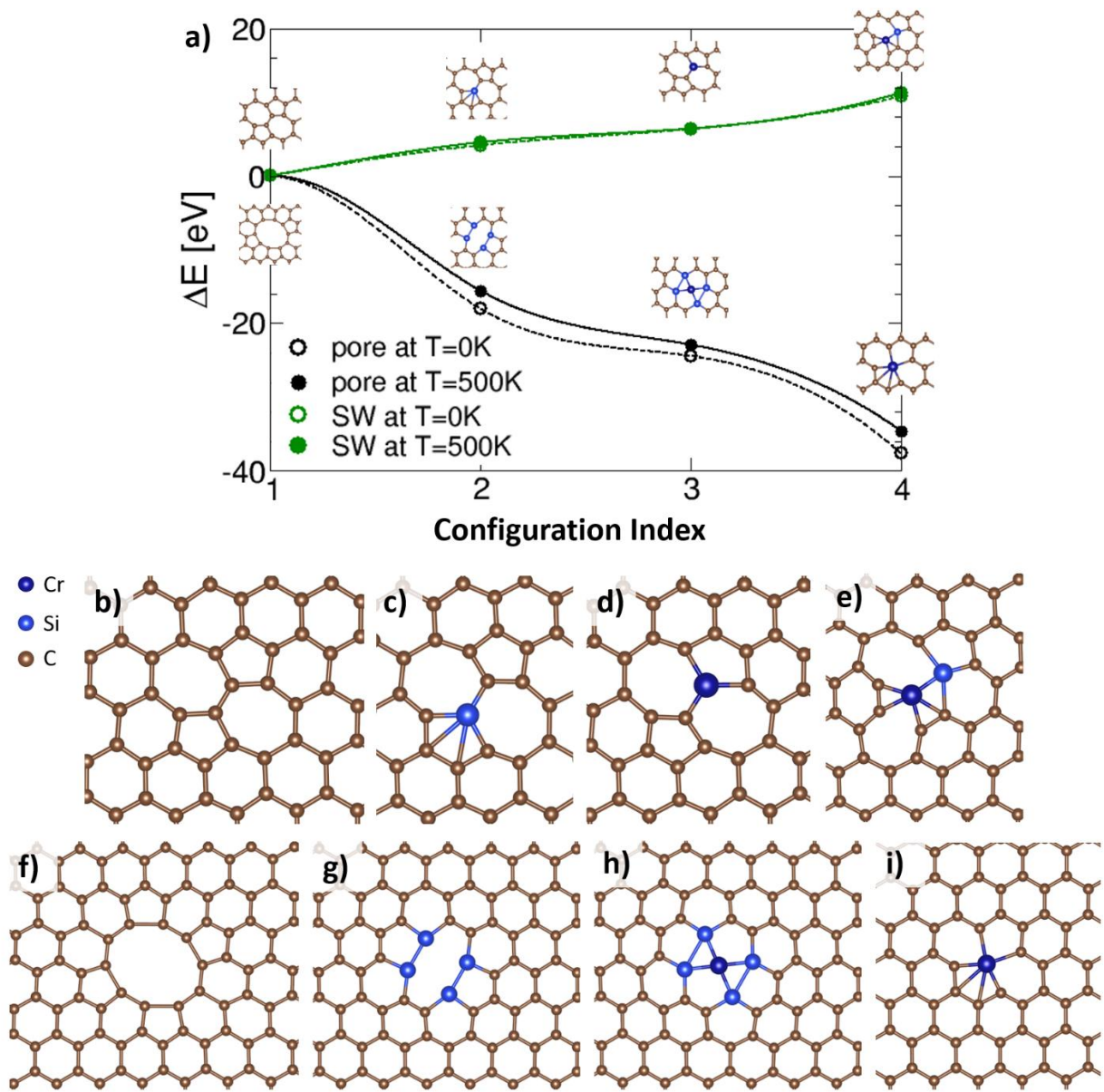

**Figure 6 Formation enthalpies of graphene defect structures containing Si or Cr atoms at Stone-Wales defects (SW) or atomic vacancies (pore) at T=0 and 500K.** Here we use graphene and gas species of Si and Cr, $Si_2$ and Cr as reference systems. a) Change in energy of each structure with respect to the starting configuration. Thumbnails of each structure are overlaid. b)-e) Magnified views of configurations evolving from a SW defect. f)-i) Magnified views of configurations evolving from a nanopore.

The total energy to incorporate dopants into the SW defect is an endothermic process; as the C atoms are replaced by dopants with a much larger radius, the energy cost increases with atomic size resulting in larger lattice distortion. By contrast, starting with a graphene nanopore, there is a significant energetic favorability for the attachment of passivating atoms along the edge. We found the effect of temperature (by repeating the calculation at T=500K) is not significant in determining the overall energetics of the metal incorporation processes. A sharp gain in energy occurs as the vacancies are terminated by four Si dopants and the cost decreases further as the unsaturated bonds are terminated by the Cr dopants and the bonds re-arrange.

It should also be noted that the formation energy of a SW defect is 4.4 eV while the formation energy of the nanopore is 44.0 eV compared to pristine graphene. So, while Figure 6 might initially suggest that the most likely evolution will follow the decreasing energetic pathway described by the nanopore passivation, this is not necessarily the case since creating the nanopore requires a significant energy input. The full process will depend on many factors, such as how the graphene is irradiated before (and during) the insertion and what types of defects are created. However, with that in mind, creating a nanopore that is then able to incorporate dopants appears to be a viable process both experimentally and energetically.

This work represents an important early step toward individual defect engineering. Future work will aim at building a more comprehensive cataloging and categorization of dopant-defect complexes and their evolution with the aid of machine learning methods as recently demonstrated[44-47] as well as detailed simulations describing the response of a material to scattering a high-energy electron. With these tools, we can begin to identify desirable functional structures as well as transition pathways toward their formation.

**Conclusion**

Creation of designer molecules and functional structures atom-by-atom remains an outstanding challenge, but if addressed, can have enormous implications for the development of new technologies with entirely new capabilities, functionalities, and applications for both classical and quantum-based technologies. Here, we demonstrate the e-beam insertion of Cr dopants into a graphene lattice *in situ* in a STEM and have shown that Cr attaches to graphene edges, sits in single, di-, and tri-vacancy defects, and interacts with co-doped Si atoms. We further revealed the energetic stability, electronic, and magnetic properties of the structures using DFT simulations. We found a significant spin polarization imbalance for a Cr atom in a graphene vacancy, which suggests a pathway for electronically controlled magnetic switching. The energetic evolution of Cr-Si heteroatomic clusters in graphene was also explored as well as whether finite temperature plays a significant role in their formation. These investigations provide an initial look at what structures are both energetically stable and experimentally realizable. This work demonstrates a route towards uncovering structures of technological interest that may be amenable to fabrication using an e-beam in a STEM with future work aimed at refining control over assembly, assembling structures with multiple atomic species, measuring properties *in situ*, and studying the role that local strain can play in tuning these properties

**Methods**

**Sample Preparation**

Chemical vapor deposition (CVD)-grown graphene was transferred to TEM grids using a wet transfer technique described in the supplemental materials. The TEM grid was baked under an Ar/O$_2$ atmosphere (90%/10%) at 500 °C for 1.5 hr to remove residual contaminants and prevent hydrocarbon deposition in the STEM.[48] Cr was then evaporated onto the graphene surface using an e-beam evaporator described more fully in the supplemental materials.

**STEM Imaging**

Imaging was conducted using a Nion UltraSTEM US200 operated at 100 kV accelerating voltage with a nominal beam current of 20 pA. The images were acquired using the medium angle annular dark field (MAADF) detector. Electron energy loss spectroscopy was performed using a Gatan Enfinium spectrometer. The nominal convergence and collection angles were respectively 30 mrad and 33 mrad.

**First-principles density functional theory calculations**

All the calculations are based on first-principles density functional theory (DFT) as implemented in the Vienna ab initio simulation package (VASP)[49] with the projector augmented wave method; a generalized gradient approximation in the form of Perdew–Burke–Ernzerhof is adopted for the exchange-correlation functional,[50] the energy cutoff of the plane-wave basis sets is 500 eV, and a $k$-mesh with $k$-spacing of $\sim 2\pi \times 0.01 \text{Å}^{-1}$ is used for the self-consistent total energy calculation. In all the calculations, a 10 Å vacuum layer is used and all atoms are fully relaxed until the residual forces on each atom are less than 0.01 eV/Å.

**Acknowledgement**

This material is based upon work supported by the U.S. Department of Energy, Office of Science, Basic Energy Sciences, Materials Sciences and Engineering Division (O.D., M.Y., A.R.L., S.J.)

**Supporting Information**

Chemical vapor deposition (CVD) was used to grow graphene on Cu foil, followed by spin coating with poly(methyl methacrylate) (PMMA) to protect the graphene. A wet graphene transfer procedure was performed as follows: The Cu/graphene/PMMA stack was set floating on the surface of an ammonium persulfate – deionized (DI) water solution (2 g ammonium persulfate in 40 ml of DI water) to dissolve the Cu foil. After the Cu was dissolved the graphene/PMMA stack was transferred to a DI water bath to rinse off any residual ammonium persulfate. The graphene/PMMA stack was then scooped from the DI water bath using a TEM grid and set to dry on a hot plate for 15 minutes at 150 °C. The grid was then immersed in acetone to dissolve the PMMA and finally dipped in isopropyl alcohol to remove the acetone and allowed to dry. The graphene samples were then baked at 500 °C in an Ar-O (90%-10%) atmosphere for 1.5 hours to remove residual contamination and prevent hydrocarbon deposition during imaging.[33] After baking, the sample was transferred to an e-beam deposition system. Metal deposition was achieved with chromium (Cr) pieces from Kurt Lesker in a Thermionics VE-240 e-beam evaporator at a pressure <5 x 10-6 Torr and at a deposition rate of~ 0.1 Å/second for a total of 3 Å. The TEM grid was fixed via Kapton tape to a plate orthogonal to the source and a shutter was used to precisely control deposition in conjunction with a quartz crystal microbalance. Just prior to examination in the STEM the sample was baked for 10 hours at 160 °C under vacuum.

Figure S7a)-c) show high angle annular dark field (HAADF) images of the suspended graphene sample at a variety of magnifications. Significant coverage of the graphene surface by chromium oxide nanoparticles was achieved. A series of electron energy loss spectra (EELS) were captured while scanning the beam over a nanoparticle, Figure S7d). The evolution of the O K edge and Cr $L_{23}$ edge with time is documented, revealing a depletion of O from the nanoparticle. Figure S7e) shows

the composition change as a function of time (lower axis) and electron dose (upper axis). The initial composition is consistent with naturally occurring $Cr_2O_3$ however the nanoparticles do not appear to have a well-ordered long-range crystalline structure. Evaluating the areal density of the nanoparticle through time reveals an increase in Cr density during the loss of O, Figure S7f), indicating that only a minimal number of Cr atoms are being sputtered. Finally, the $L_3/L_2$ ratio was measured, using a double arctangent background subtraction model[34] and integration window widths of 7.2 eV, which revealed a decrease from ~1.8 to ~1.3, also consistent with conversion from $Cr_2O_3$ to Cr.[35] Since the loss of O was completed during the first ~450 s this gives an opportunity to evaluate the stability of the $L_3/L_2$ measurement. After 450 s we can see that the $L_3/L_2$ ratio varies by about $\pm 0.1$ and that the prior decrease was significantly larger than this fluctuation. Thus, the preferential sputtering of O from the $Cr_2O_3$ nanoparticles provides a convenient mechanism for us to isolate Cr using the e-beam. Cr atoms were also occasionally observed to sputter from the nanoparticles under the influence of the e-beam and could be seen in the lighter contamination between the nanoparticles.

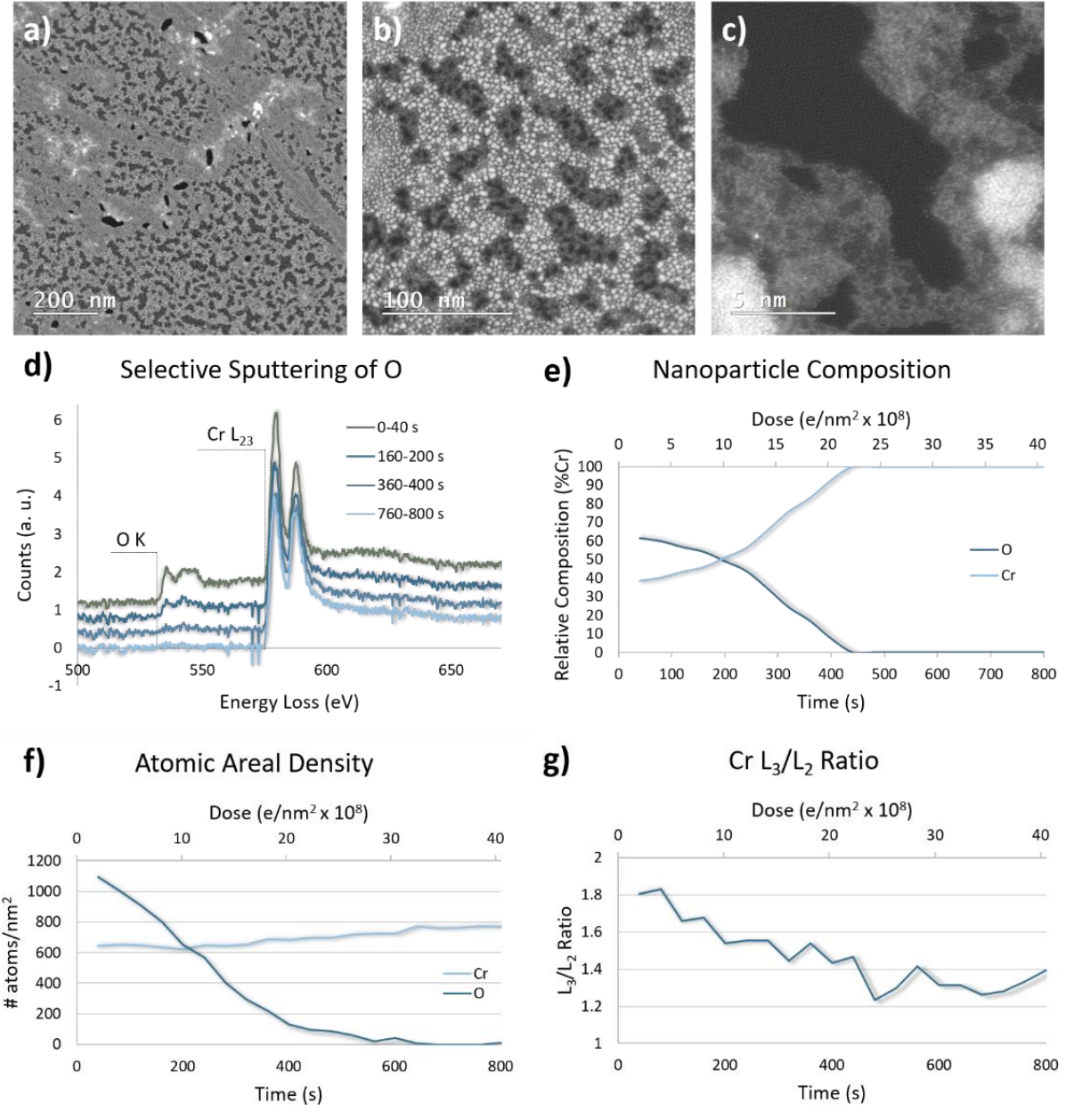

**Figure S7 Overview of Cr/graphene sample and characterization of the electron beam effect on Cr$_2$O$_3$ nanoparticles.** a)-c) Various magnification HAADF-STEM images of suspended graphene which has been coated with Cr$_2$O$_3$ nanoparticles through *ex situ* e-beam evaporation. d) EELS spectra acquired through time while scanning the electron beam over a Cr$_2$O$_3$ nanoparticle. We observe a steady decrease in the O K edge intensity indicating a selective sputtering of oxygen. e) The relative composition of the nanoparticle (vertical axis) is shown as a function of electron dose (top axis) and time (bottom axis). f) The atomic areal density of Cr and O is shown as a function of time (bottom axis) and electron dose (top axis). We observe an increase in Cr density with removal of O, consistent with conversion from $Cr_2O_3$ to Cr. g) The Cr $L_3/L_2$ ratio is shown as a function of time (bottom axis) and electron dose (top axis) and is consistent with the conversion of trivalent $Cr_2O_3$ to zero-valent elemental Cr.

Examining the material between the nanoparticles we find a mixture of Cr, C, and Si. The brightest of these is expected to be Cr, based on the contrast mechanism of this imaging mode.[36]

**Verification of Cr atoms using EELS**

To verify that the bright atoms dispersed in the surface contamination were Cr atoms, an EELS image was acquired over an area of the sample boxed in Figure S8a). The simultaneously acquired HAADF image is shown in the inset. Because the Cr atoms are mobile we were unable to position the stationary beam on them, however, in the spectrum image we can capture spectra over a larger area and then look for the position of the bright atoms post acquisition. We can see that several bright atoms were captured in the spectrum image (though not in their original positions). Selecting this portion of the spectrum image we observe an unambiguous Cr EELS signal which confirms the bright atoms are Cr. Figure S8b shows the spectrum from the location boxed in the spectrum image with a clear Cr $L_{23}$ edge and no O K edge. Figure S8c,d shows a similar analysis but capturing the Si $L_{23}$ and C K edge. This is consistent with previous observations of the graphene contamination being primarily composed of Si and C.

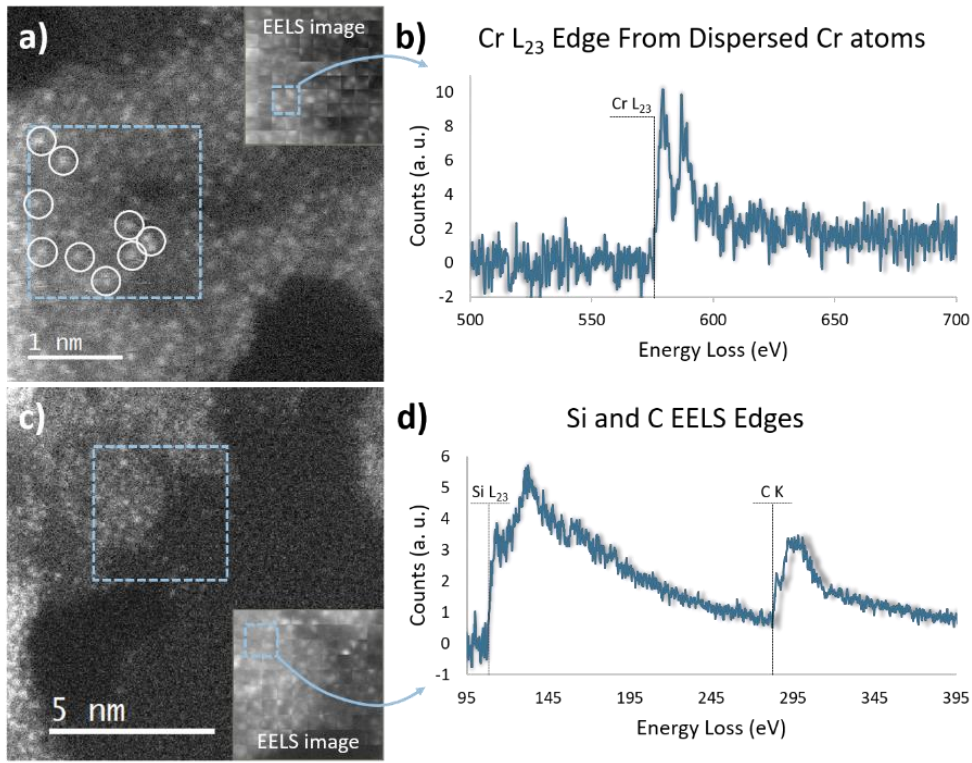

**Figure S8 Characterization of the material found near the nanoparticles adjacent to pristine areas of graphene.** a) Overview HAADF-STEM image of the material showing atoms of different intensities suggesting the presence of different elements. Several brighter atoms are circled. Since the atoms were seen to move quite freely during observation an EELS image was acquired over the dotted area. The inset shows the HAADF-STEM image captured during the EELS acquisition. Four pixels were summed together (indicated by the dotted box in the inset) to extract the EELS signal, b), from single Cr atoms moving within the material. c) HAADF-STEM overview image of a similar area. The dotted box indicates the region an EELS spectrum image was acquired. Inset is the HAADF-STEM image captured during the EELS acquisition. Four pixels were summed together (indicated by the dotted box in the inset) to extract the EELS signal, d), of the Si and C edges. This data suggests the material is primarily an amorphous mixture of Si and C with a few Cr atoms scattered around.

**Additional example of beam dragging Cr atoms into the graphene lattice**

The e-beam was used to sequentially damage the graphene lattice and scatter nearby Cr atoms onto the defect locations. Figure S9 illustrates the progression of this process through time. Figure S9a shows the initial configuration where a clean patch of suspended graphene is adjacent to a contaminated region with some bright Cr atoms. The e-beam was then manually positioned in the circled regions in Figure S9a-c, where the Cr atoms were located, and dragged in the direction indicated by the arrows. We observed that the Cr atoms could be moved along the edge of the bilayer area and out onto the single layer graphene. The highly focused, 100 kV, beam is sufficient to quickly eject C atoms from the graphene lattice allowing the Cr atoms to bond at the defect sites and stabilize (e.g. long enough to capture images even with the 100 kV e-beam). The final configuration of the sample is shown in Figure S9d where seven Cr atoms have been attached to the graphene lattice. In Figure S9c and d we also note that a few Si atoms have also been attached, characterized by their dimmer contrast.

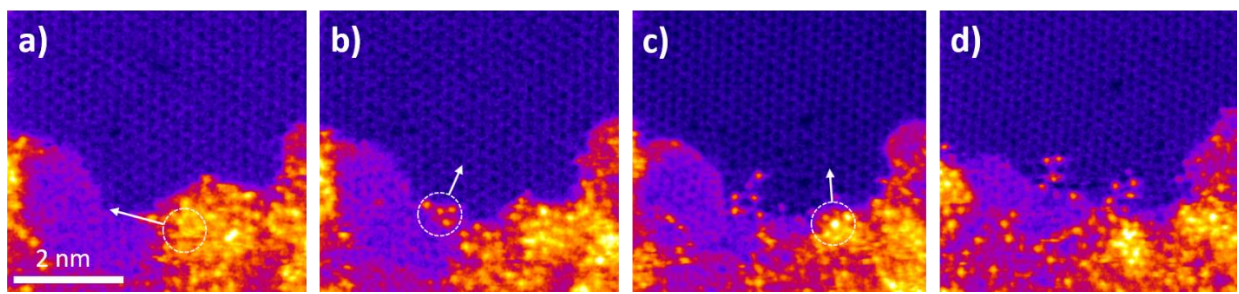

**Figure S9 A second example of using the e-beam to attach Cr atoms to graphene defects.** The initial configuration is show in a). The e-beam was positioned within the marked circle and dragged across the sample in the direction indicated by the arrow. In b) we see that a few Cr atoms have been separated from the source material and reside at the edge of the pristine graphene. The beam was again dragged across the sample as indicated by the arrow. c) shows the result of this procedure where a few Cr atoms now lie on top of the graphene lattice or have been attached to defect sites. The beam dragging process was repeated to produce the final configuration shown in d).

**Convergence test of DFT results**

Figure S4 summarizes the total energy and magnetic moment of 5x5 Cr-doped divancy system with the number of k-points and the energy cutoff. The magnetic moment and total energy are converged with k-mesh >6x6x1 and energy cut off > 350eV.

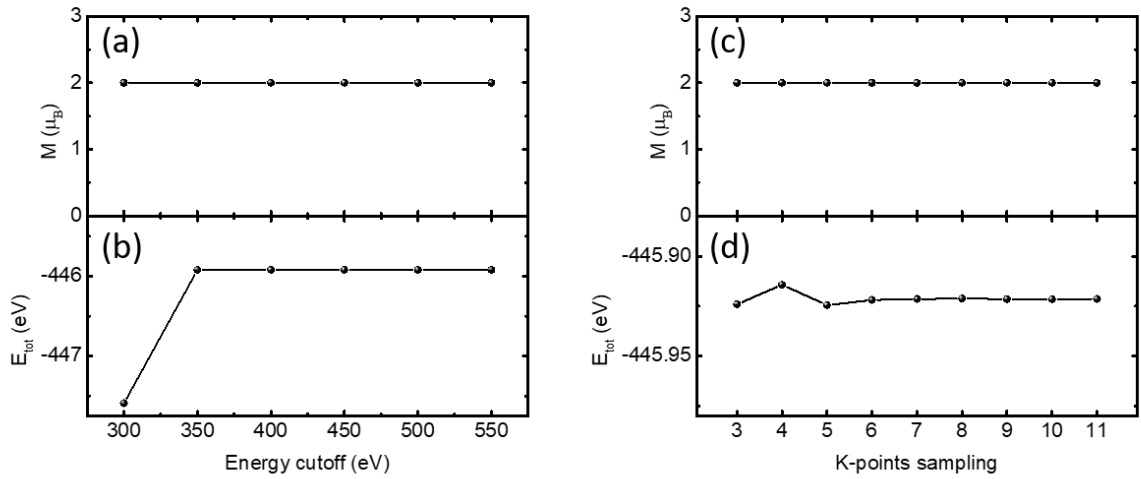

**Figure S4.** Magnetic moment and total energy of 5x5 Cr-doped at divacancy (DV) in terms of energy cutoff [(a) and (b)] and k-mesh [(c) and (d)], where for (a) and (b), the k-mesh of 9X9X1, and for (c) and (d), the energy cutoff of 400 eV are employed.

In the following we compare the projected density of states (PDOS) of three different model systems (monovacancy (MV), divancy (DV), and trivacancy (TV)), where Cr atoms adsorbed at the vacancy sites to form "flat" or "corrugated" configurations. In addition, we choose PBE+U approach to systematically analyze the effect of electron correlation of Cr d orbitals in the overall properties. The advantage of PBE+U over hybrid functional including the portion of the Hartree-Fock term is that we can pinpoint the impact of the specific orbital that may include strong electron correlation. The U value of Cr d-orbital was 2.5eV.

PDOS of Figure 4, using the flat configuration using LDA functional, is compared with the result using PBE/PBE+U in Fig. S5. We evaluate the PDOS of the corrugated configurations using PBE and PBE+U in Fig. S6. For the given configuration, the introduction of the electron correlation in the form of PBE+U, doesn't significantly affect the overall properties. We thus conclude that the Coulomb correlation energy only influences the energy gaps of these three models. On the other hand, the local corrugation of Cr atoms creates a noticeable influence in the projected density of states, specifically the size of the gaps is changing greatly with these factors.

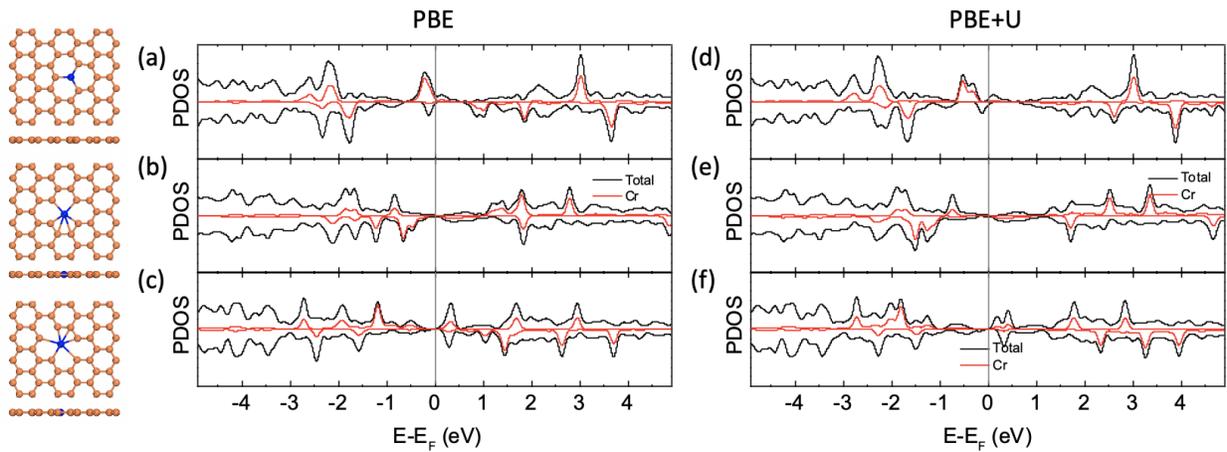

**Figure S5.** PDOS of "flat" configuration with three different vacancy structures, Cr@MV, Cr@DV, and Cr@TV, using PBE functional (a-c) and PBE+U functional (d-f).

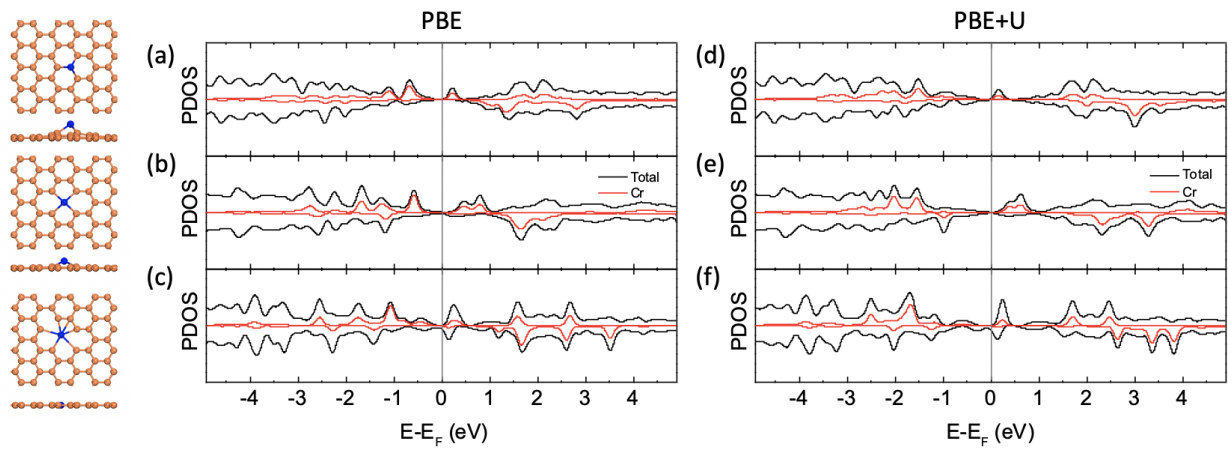

**Figure S6.** PDOS of "corrugated" configuration with three different vacancy structures, Cr@MV, Cr@DV, and Cr@TV models with PBE functional (a-c) and PBE+U functional (d-f).